\documentclass[aps,prc,twocolumn,superscriptaddress,showpacs,floatfix,nofootinbib]{revtex4-1}
\usepackage{amsmath,graphicx,color}
\begin{document}
\title{
Effective shear and bulk viscosities for anisotropic flow}
\author{Fernando G. Gardim}
\affiliation{Instituto de Ci\^encia e Tecnologia, Universidade Federal de Alfenas, 37715-400 Po\c cos de Caldas, MG, Brazil}
 \author{Jean-Yves Ollitrault}
\affiliation{Université Paris-Saclay, CNRS, CEA, Institut de physique th\'eorique, 91191, Gif-sur-Yvette, France} 
\date{\today}

\begin{abstract}
We evaluate the viscous damping of anisotropic flow in heavy-ion collisions for arbitrary temperature-dependent shear and bulk viscosities. 
We show that the damping is solely determined by effective shear and bulk viscosities, which are weighted averages over the temperature. 
We determine the relevant weights for nucleus-nucleus collisions at $\sqrt{s_{\rm NN}}=5.02$~TeV and 200~GeV, corresponding to the maximum LHC and RHIC energies, by running ideal and viscous hydrodynamic simulations. 
The effective shear viscosity is driven by temperatures below $210$~MeV at RHIC, and below $280$~MeV at the LHC, with the largest contributions coming from the lowest temperatures, just above freeze-out. 
The effective bulk viscosity is driven by somewhat higher temperatures, corresponding to earlier stages of the collision. 
We show that at a fixed collision energy, the effective viscosity is independent of centrality and system size, to the same extent as the mean transverse momentum of outgoing hadrons. 
The variation of viscous damping is determined by Reynolds number scaling. 
\end{abstract}
\maketitle

\section{Introduction}
Determining the transport coefficients of the quark-gluon plasma, such as its shear ($\eta$) and bulk ($\zeta$) viscosities, is one of the goals of heavy-ion physics. 
One of the motivations is the early recognition that the quark-gluon plasma produced in heavy-ion collisions has a very low shear viscosity over entropy ($\eta/s$) ratio~\cite{Romatschke:2007mq}, implying the formation of a strongly-coupled fluid~\cite{Kovtun:2004de}.
Shear viscosity is now included in the vast majority of state-of-the art hydrodynamic simulations of heavy-ion collisions~\cite{Heinz:2013th}. 
It has been shown that bulk viscosity must also be taken into account in order to quantitatively explain experimental data~\cite{Ryu:2015vwa}.

Theoretical calculations of transport coefficients are notoriously difficult. 
Perturbative results~\cite{Arnold:2003zc,Ghiglieri:2018dib} are accurate only at temperatures much higher than those achieved in the laboratory. 
Ab-initio calculations of transport coefficients with lattice techniques pose serious numerical and theoretical challenges~\cite{Bazavov:2019lgz}.
Results have been obtained only for the gluon plasma without quarks, both for shear viscosity~\cite{Meyer:2007ic,Astrakhantsev:2017nrs} and for bulk viscosity~\cite{Meyer:2007dy,Astrakhantsev:2018oue}. 
To include quarks, one must resort to functional techniques~\cite{Christiansen:2014ypa} or effective models~\cite{Mykhaylova:2019wci}. 
There is a theoretical consensus that both $\eta/s$ and $\zeta/s$ depend strongly on temperature ($\eta/s$ typically increases as a function of $T$ above 160~MeV, while $\zeta/s$ decreases). 
Over the last decade, several efforts have been made to incorporate this temperature dependence into hydrodynamic calculations~\cite{Denicol:2010tr,Niemi:2011ix,Niemi:2012ry,Dubla:2018czx}.

An important question is how the temperature dependence of $\eta/s$ and $\zeta/s$ can be constrained using experimental data~\cite{Niemi:2015qia,Bernhard:2019bmu}.
A recent study shows that $\eta/s$ is most constrained in the temperature range $T\sim 150-220$~MeV~\cite{Auvinen:2020mpc}. 
The phenomenon that allows one to best constrain transport coefficients is anisotropic flow~\cite{Heinz:2013th}, by which the distribution of outgoing particles breaks azimuthal symmetry.
The azimuthal anisotropy, which is characterized by Fourier coefficients $v_n$, builds up gradually as a result of the collective expansion~\cite{Ollitrault:2007du}.
Viscosity makes the expansion less collective, thus reducing $v_n$. 

We carry out a systematic investigation of this decrease for the two largest harmonics, $v_2$~\cite{Romatschke:2007mq} and $v_3$~\cite{Alver:2010dn}. 
Our study is limited to the integrated anisotropic flow (averaged over transverse momenta) because, as will be argued below, its determination in viscous hydrodynamics is more robust than that of differential flow.  
In Sec.~\ref{s:linearity}, we show that the reduction in $v_n$ due to viscosity can be written as a weighted integral of the temperature-dependent $\eta/s$ and $\zeta/s$. 
We define effective viscosities, which encapsulate the information on viscosity that one can gain from anisotropic flow. 
In Sec.~\ref{s:weights}, we determine the weights that define the effective viscosities by running hydrodynamic simulations of central Pb+Pb collisions at $\sqrt{s_{\rm NN}}=5.02$~TeV.
In Sec.~\ref{s:dimensional}, we check that the order of magnitude of viscous damping is compatible with expectations from dimensional analysis. 
In Sec.~\ref{s:linearitytests}, we show that the effective viscosity is an excellent predictor of the viscous suppression of $v_n$ for a wide range of temperature-dependent shear and bulk viscosities. 
In Sec.~\ref{s:centrality}, we check that the centrality and system-size dependence of the viscous damping follows the $1/R$ scaling expected from dimensional analysis, where $R$ is the transverse size. 
The dependence on collision energy is illustrated in Sec.~\ref{s:rhic} where we carry out calculations at $\sqrt{s_{\rm NN}}=200$~GeV, corresponding to the top RHIC energy. 

\section{Effective viscosity}
\label{s:linearity}

We define effective bulk and shear viscosities of hot quark and gluon matter,  which determine the damping of anisotropic flow. 

A hydrodynamic simulation starts from an initial condition, corresponding typically to the entropy density profile at an early time. 
One then solves the equations of hydrodynamics, which model the expansion of the system into the vacuum. 
We study the effect of viscosity by evolving the same initial profile through ideal hydrodynamics ($\eta/s=\zeta/s=0$) and viscous hydrodynamics.
The fluid eventually fragments into individual hadrons, and we evaluate $v_n$ from the distribution of outgoing particles in both cases. 
We use the following quantity as a measure of the viscous damping: 
\begin{equation}
\label{defDelta}
  \Delta_n\equiv\ln\left( \frac{v_n({\rm viscous})}{v_n({\rm ideal})}\right).
\end{equation}
If $|\Delta_n|\ll 1$, then, $\Delta_n$ is the relative change of $v_n$ due to viscosity, $\Delta_n\simeq v_n({\rm viscous})/v_n({\rm ideal})-1$.
One typically expects viscosity to reduce $v_n$~\cite{Romatschke:2007mq}, resulting in a negative $\Delta_n$.

Our study is limited to $v_2$ and $v_3$ because their dependence on the initial density profile is, to a good approximation~\cite{Gardim:2014tya}, a linear response to the corresponding initial anisotropy $\varepsilon_n$~\cite{Teaney:2010vd}, both in ideal and viscous hydrodynamics, so that the dependence cancels when taking the ratio in Eq.~(\ref{defDelta}). 
Therefore, even though we evaluate $\Delta_n$ with a specific, smooth density profile, which will be specified in Sec.~\ref{s:weights}, we expect the result to be universal to a good approximation.
This should however be checked explicitly when initial-state fluctuations are present~\cite{Aguiar:2001ac,Miller:2003kd,Alver:2006wh}.
We plan to do this in a future work. 

We now derive the general expression of $\Delta_n$ in the limit of small viscosities. 
$\Delta_n$ is a functional of $(\eta/s)(T)$ and $(\zeta/s)(T)$, which vanishes by construction if $(\eta/s)(T)=(\zeta/s)(T)=0$.
Transport coefficients enter the relativistic Navier-Stokes equations (which represent the first-order correction due to viscosity) as two separate linear contributions~\cite{Romatschke:2017ejr}. 
Therefore, for small $(\eta/s)(T)$ and $(\zeta/s)(T)$, $\Delta_n$ must be a linear functional of these quantities~\cite{Paquet:2019npk}:
\begin{equation}
\label{convolution}
\Delta_n=\int_{T_f}^\infty \frac{\eta}{s}(T)w^{(\eta)}_n(T)dT+\int_{T_f}^\infty \frac{\zeta}{s}(T)w^{(\zeta)}_n(T)dT,
\end{equation}
where $T_f$ is the lowest value of the temperature, called the freeze-out temperature, and $w^{(\eta)}_n(T)$ and $w^{(\zeta)}_n(T)$ are weight functions for shear and bulk viscosity.
These weight functions quantify the effect of viscosity on anisotropic flow at a given temperature. 

Note that the separation of shear and bulk viscosity in Eq.~(\ref{convolution}) strictly holds only to first order in $\eta$ and $\zeta$. 
To second order, one expects on purely mathematical grounds an additional term coupling shear and bulk in the form of a double integral
$\int \eta(T_1)\zeta(T_2) f(T_1,T_2)dT_1dT_2$. 
Such a coupling is naturally expected in second-order viscous hydrodynamics, where shear and bulk viscosities are intertwined via their relaxation equations (Eqs.~(3) and (4) of \cite{Paquet:2015lta}). 
It is not studied in this paper. 

We define the effective shear and bulk viscosities relevant for $v_n$ by: 
\begin{eqnarray}
  \label{defeffective}
  \left(\frac{\eta}{s}\right)_{n,\rm eff}&=&\frac{\int_{T_f}^\infty (\eta/s)(T)w^{(\eta)}_n(T)dT}{\int_{T_f}^\infty w^{(\eta)}_n(T)dT}\cr
  \left(\frac{\zeta}{s}\right)_{n,\rm eff}&=&\frac{\int_{T_f}^\infty (\zeta/s)(T)w^{(\zeta)}_n(T)dT}{\int_{T_f}^\infty w^{(\zeta)}_n(T)dT}.
\end{eqnarray}
Then, Eq.~(\ref{convolution}) expresses the damping of $v_n$ as 
\begin{equation}
\label{predicted}
\Delta_{n}
=W^{(\eta)}_n\times\left(\frac{\eta}{s}\right)_{n,\rm eff}+W^{(\zeta)}_n\times\left(\frac{\zeta}{s}\right)_{n,\rm eff},
\end{equation}
where
\begin{equation}
  \label{integral}
W^{(\eta,\zeta)}_n\equiv \int_{T_f}^\infty w^{(\eta,\zeta)}_n(T)dT.
\end{equation}
Equation~(\ref{predicted}) implies that for any temperature-dependent viscosity, the damping of $v_n$ is solely determined by the effective shear and bulk viscosities defined by Eq.~(\ref{defeffective}). 
This result holds in the limit of small viscosity.
Note, however, that the validity of hydrodynamics itself requires that viscosity has a small relative effect on observables, since viscous hydrodynamics is the first term in a systematic gradient expansion~\cite{Baier:2007ix}. 
We therefore postulate that our result is general, and that the damping of $v_n$ is always determined by the effective viscosities. 
This will be checked explicitly in Sec.~\ref{s:linearitytests}. 

The effective viscosity (\ref{defeffective}) is a weighted average of the temperature-dependent viscosity. 
It is similar to the quantity recently introduced by Paquet {\it et al.\/}~\cite{Paquet:2019npk}, but applied to different observables (anisotropic flow, as opposed to entropy), so that weights are different. 
We determine the relevant weights for the integrated anisotropic flow in Sec.~\ref{s:weights}.
We then test the validity of Eq.~(\ref{predicted}) in Sec.~\ref{s:linearitytests}. 

\begin{figure*}[ht]
\begin{center}
\includegraphics[width=.8\linewidth]{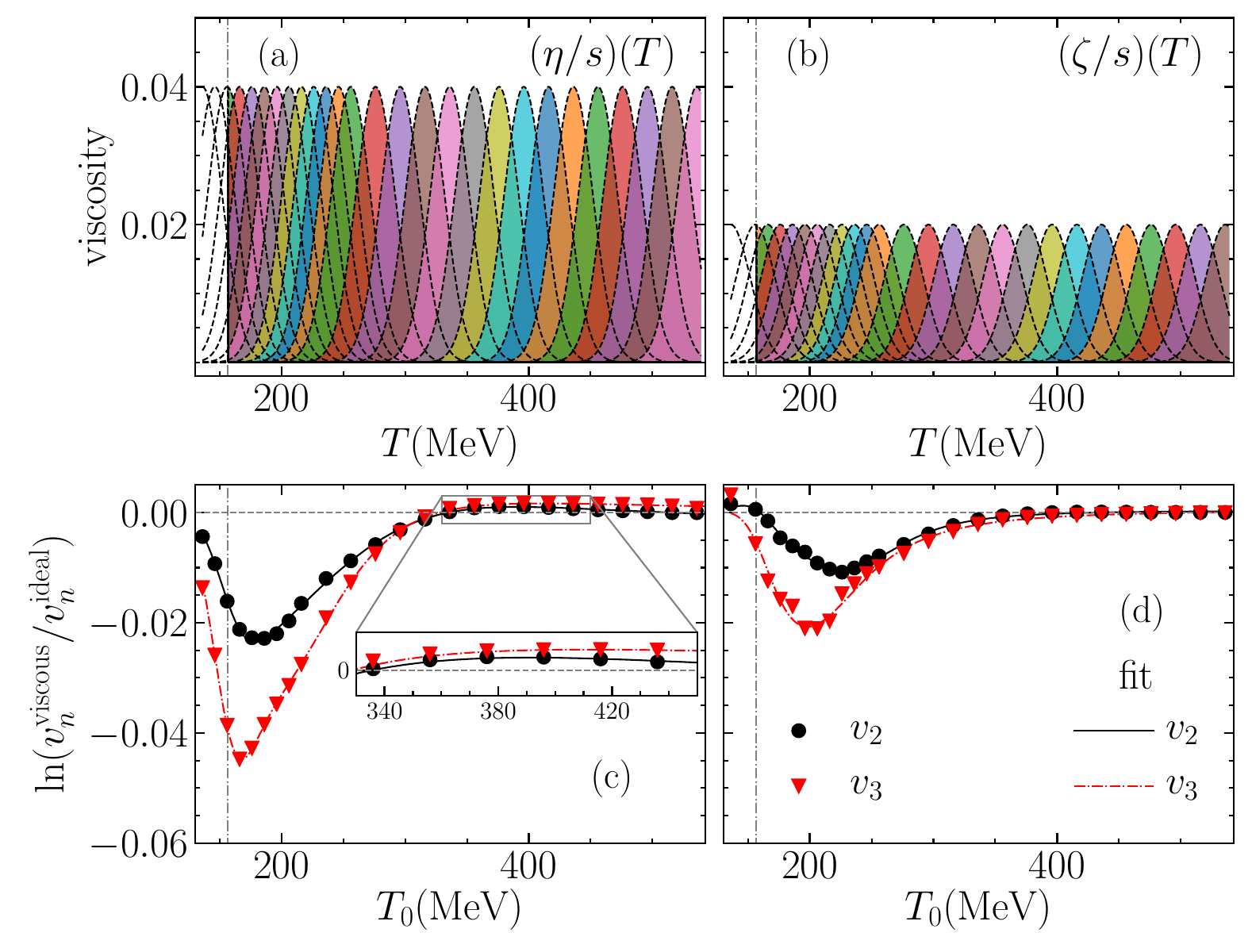} 
\end{center}
\caption{(Color online)
  Effect of a temperature-dependent shear ($\eta$) or bulk ($\zeta$) viscosity on $v_2$ and $v_3$ in central Pb+Pb collisions at $\sqrt{s_{\rm NN}}=5.02$~TeV. 
  Shear corresponds to the left panels (a) and (c), bulk to the right panels (b) and (d). 
  The upper panels display the $(\eta/s)$  and $(\zeta/s)$ profiles used in our caculation.
  They are defined by Eq.~(\ref{peak}), with $\sigma=16$~MeV, $(\eta/s)_{\max}=0.04$ (a) and $(\zeta/s)_{\max}=0.02$ (b), and each curve corresponds to a different value of $T_0$. 
The vertical lines indicate the freeze-out temperature $T_f=156.5$~MeV. 
The symbols in panels (c) and (d) display the corresponding values of $\Delta_n$, defined by Eq.~(\ref{defDelta}), as a function of $T_0$. 
Lines are fits using Eqs.~(\ref{convolution}) and (\ref{deltaplussmooth}). 
}
\label{fig:delta}
\end{figure*}    

\section{Determining the weighting functions}
\label{s:weights}

In this Section, we determine the weighting functions $w_n^{(\eta)}(T)$ and $w_n^{(\zeta)}(T)$, which define the effective viscosity (\ref{defeffective}), for central Pb+Pb collisions at the top LHC energy $\sqrt{s_{\rm NN}}=5.02$~TeV.
We carry out two separate sets of hydrodynamic simulations, one with only shear viscosity and one with only bulk viscosity. 
In order to isolate the effect of the viscosity in a specific temperature range, we implement a viscosity profile which is a narrow window of width $\sigma$, centered around a temperature $T_0$:
\begin{equation}
\label{peak}
\frac{\eta}{s}(T)=\left(\frac{\eta}{s}\right)_{\max}\exp\left(-\frac{(T-T_0)^2}{2\sigma^2}\right),
\end{equation}
where $(\eta/s)_{\max}$ the maximum value of $\eta/s$. 
We carry out simulations for a large number of values of $T_0$, which span the range of temperatures in a heavy-ion collision.
The exact same procedure is repeated for bulk viscosity, replacing $\eta/s$ with $\zeta/s$.

The first thought would be to use a window as narrow as possible. 
If $\sigma$ is too small, however, there are large errors for the following reason: 
The viscosity varies steeply with the temperature, which itself depends on space-time coordinates.
This results in large pressure gradients, while they should always be small in hydrodynamics~\cite{Baier:2007ix}.
These gradients are proportional to $(\eta/s)_{\max}/\sigma$.
When gradients are too large, we find that instabilities occur, which appear as numerical errors (e.g., $v_n$ jumping up and down upon small variations of $T_0$). 
We have adjusted the values of parameters so that results are stable. 
Our simulations are carried out with $\sigma=16$~MeV, $(\eta/s)_{\max}=0.04$ (Fig.~\ref{fig:delta} (a)) and $(\zeta/s)_{\max}=0.02$ (Fig.~\ref{fig:delta} (b)).

\begin{figure}[ht]
\begin{center}
\includegraphics[width=\linewidth]{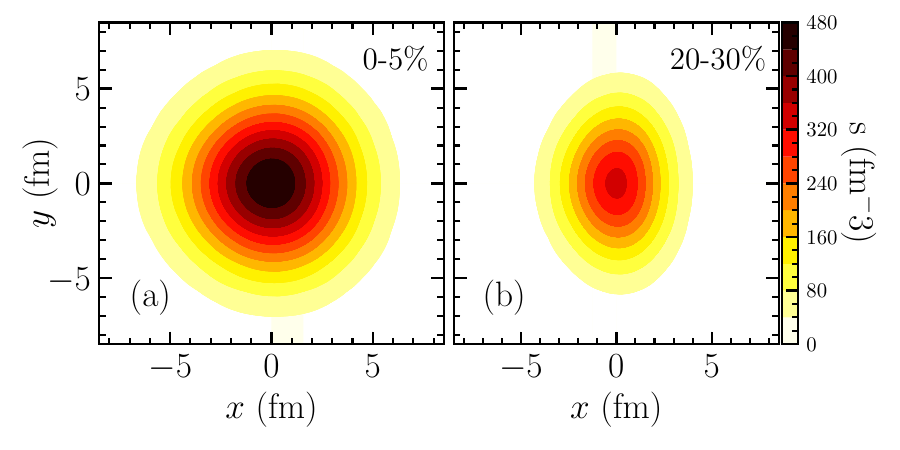} 
\end{center}
\caption{(Color online)
  Initial entropy density profile used in our ideal hydrodynamic calculation, defined by Eq.~(\ref{initial}), with parameters tuned to match Pb+Pb collisions at $\sqrt{s_{\rm NN}}=5.02$~TeV. 
(a) 0-5\% centrality window (Secs.~\ref{s:weights}---\ref{s:linearitytests}): $s_0=438$~fm$^{-3}$, $R_0=4.18$~fm, $\varepsilon_2 = 0.085$, $\varepsilon_3=0.075$. 
  (b) 20-30\% centrality window (Sec.~\ref{s:centrality}): $s_0=337$~fm$^{-3}$, $R_0=2.97$~fm, $\varepsilon_2 =0.35$, $\varepsilon_3=0.12$.
  The profile is identical for the viscous hydrodynamic calculation, except for the overall normalization (see text). 
}
\label{fig:density}
\end{figure}    

Our hydrodynamic simulation uses boost-invariant~\cite{Bjorken:1982qr} initial conditions, with a starting time $\tau_0=0.6$~fm/c.
The transverse velocity at $\tau_0$ is set to zero, that is, initial flow~\cite{Vredevoogd:2008id,vanderSchee:2013pia} is neglected.
The initial entropy density profile is a deformed Gaussian~\cite{Alver:2010dn}:
\begin{eqnarray}
  \label{initial}
  s(x,y)&=&  s(r\cos\phi,r\sin\phi)\cr
  &=&s_0
  \exp\left(-\frac{r^2}{R_0^2}\left(1+\varepsilon_2\cos 2\phi+\frac{4}{5}\varepsilon_3\cos 3\phi\right)\right). 
\end{eqnarray}
In this equation, $\varepsilon_2$ and $\varepsilon_3$ are the initial eccentricities~\cite{Petersen:2010cw},\footnote{$\varepsilon_2$ and $\varepsilon_3$ in Eq.~(\ref{initial}) correspond to the usual eccentricities~\cite{Yan:2017bgc} only in the limit where they are much smaller than unity, more precisely, to first order in $\varepsilon_2$ and $\varepsilon_3$.} which produce elliptic flow and triangular flow after hydrodynamic expansion. 

We run ideal and viscous hydrodynamics with the same values of $R_0$, $\varepsilon_2$ and $\varepsilon_3$, not with the same normalization $s_0$. 
The reason is that the normalization is typ!cally adjusted so as to match the observed multiplicity. 
It seems natural to compare ideal and viscous hydrodynamics at the same multiplicity. 
Now, the multiplicity is proportional to the final entropy, which is larger than the initial entropy in viscous hydrodynamics~\cite{Hanus:2019fnc}. 
Therefore, the normalization $s_0$ must be lowered in viscous hydrodynamics.\footnote{In practice, we choose $s_0$ for viscous hydrodynamics so that the
final multiplicity is close to the expected value. For efficiency, we
then estimate $v_n$ for the corresponding multiplicity in ideal
hydrodynamics from an array of ideal hydrodynamics calculations by
linear interpolation.}

We fix the parameters of Eq.~(\ref{initial}) as follows: 
We evaluate $R_0$ by matching the rms radius to a model of initial conditions that reproduces well the mean transverse momentum $\langle p_t\rangle$~\cite{Giacalone:2020dln}, since $\langle p_t\rangle$ is determined by the initial radius in ideal hydrodynamics~\cite{Broniowski:2009fm,Bozek:2017elk}.\footnote{Note that we compare viscous and ideal hydrodynamics with the same value of multiplicity and $R_0$. It could have been better to fix $\langle p_t\rangle$, rather than $R_0$. However, viscosity has a smaller relative effect on $\langle p_t\rangle$ than on $v_n$, so that the final results would likely be similar.} 
We then fix the normalization constant $s_0$ in such a way that the multiplicity matches that measured in Pb+Pb collisions at $\sqrt{s_{\rm NN}}=5.02$~TeV~\cite{Adam:2015ptt}. 
Finally, we evaluate $\varepsilon_2$ and $\varepsilon_3$ from a model of initial conditions~\cite{Gelis:2019vzt} which reproduces well the measured values of $v_2$ and $v_3$.\footnote{This is not crucial as our final results are independent of $\varepsilon_2$ and $\varepsilon_3$.}
The resulting density profile is represented in Fig.~\ref{fig:density} (a) for central collisions.  

We then evolve this initial condition using the MUSIC hydrodynamic code~\cite{Schenke:2010nt,Schenke:2011bn,Paquet:2015lta} with a realistic equation of state inspired by lattice QCD~\cite{Huovinen:2009yb}.
We evaluate $v_2$ and $v_3$ at the freeze-out temperature $T_f=156.5$~MeV~\cite{Bazavov:2018mes}.
The viscous corrections to the momentum distribution at freeze out are evaluated using the usual quadratic ansatz~\cite{Teaney:2003kp,Bozek:2009dw}.
We take into account hadronic decays, but we neglect rescatterings in the hadronic phase. 
For the sake of simplicity, we evaluate $v_2$ and $v_3$ (averaged over all transverse momenta $p_t$), in the same pseudorapidity window $|\eta|<0.5$ used to measure the multiplicity~\cite{Adam:2015ptt}.
The fact that experiments use different pseudorapidity cuts~\cite{Adam:2016izf} matters little, since these kinematic cuts typically multiply $v_n$ by a constant factor, which cancels when evaluating $\Delta_n$ using Eq.~(\ref{defDelta}). 

\begin{figure}[ht]
\begin{center}
\includegraphics[width=\linewidth]{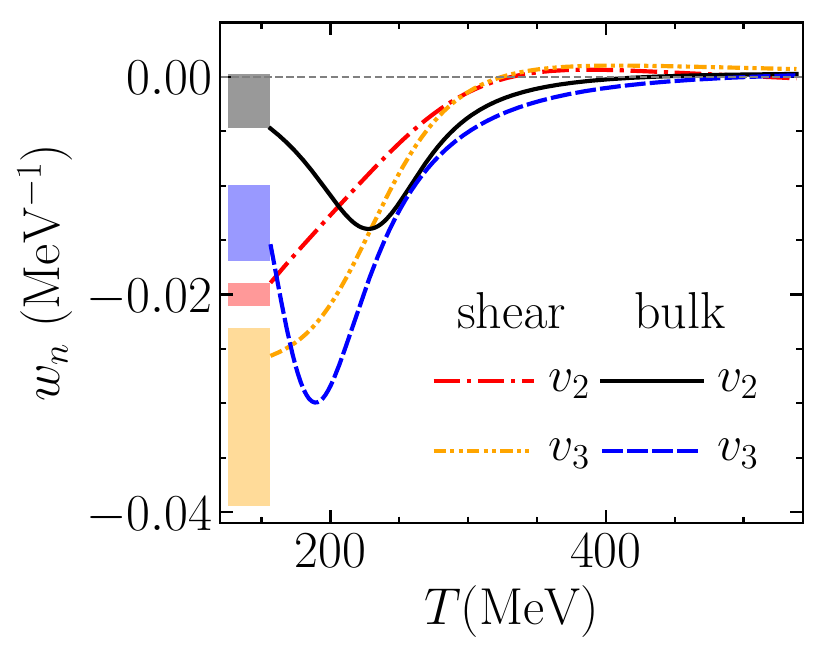} 
\end{center}
\caption{(Color online)
  Weights $w_2^{\zeta}(T)$ (full line), $w_3^{\zeta}(T)$ (dashed line), $w_2^{\eta}(T)$ (dot-dashed line), $w_3^{\eta}(T)$ (dotted line), defining the effective viscosities (\ref{defeffective}) at $\sqrt{s_{\rm NN}}=5.02$~TeV. 
  They are obtained by fitting the hydrodynamic results in Fig.~\ref{fig:delta} using Eq.~(\ref{deltaplussmooth}).
  The shaded boxes to the left are meant to represent the discrete part $w_f\delta(T-T_f)$: Their area is $|w_f|$ (see numbers in Table~\ref{tableweights}), and their vertical placement is arbitrary.
  }
\label{fig:weights}
\end{figure}    

The values of $\Delta_n$ are displayed in Fig.~\ref{fig:delta} (c) and (d) for shear and bulk viscosity, as a function of the temperature $T_0$ in Eq.~(\ref{peak}). 
$\Delta_n$ is mostly negative, which means that viscosity decreases anisotropic flow~\cite{Romatschke:2007mq}.
$\Delta_2$ and $\Delta_3$ have similar variations as a function of $T_0$, but the overall magnitude of $\Delta_3$ is larger, both for shear and bulk viscosity:
As expected, damping is stronger for higher harmonics~\cite{Alver:2010dn,Teaney:2012ke}.
Large negative values of $\Delta_n$ are obtained for values of $T_0$ around $200$~MeV, corresponding to the late stages of the hydrodynamic evolution. 
For $T_0>300$~MeV, corresponding to a viscosity which is only present during the early stages, $\Delta_n$ is much smaller. 
Interestingly, for shear viscosity, $\Delta_n$ changes sign and becomes positive for $T_0>330$~MeV (see the zoom in Fig.~\ref{fig:delta} (c)).
This implies that shear viscosity at high temperature {\it increases\/} $v_n$, although by a very modest amount. 
The physical interpretation is that when the longitudinal expansion dominates, shear viscosity reduces the longitudinal pressure and increases the transverse pressure, leading to an increased transverse flow in general, and anisotropic flow in particular.

Using the results for $\Delta_n$, we then infer $w_n^{(\eta,\zeta)}(T)$ defined by Eq.~(\ref{convolution}). 
One would naively expect $w_n^{(\eta,\zeta)}(T)$ to be a smooth function of $T$. 
However, one must remember that viscosity enters a hydrodynamic simulation in two different places:
(1) In the equations of hydrodynamics themselves; (2) At the final stage when the fluid is transformed into particles. 
The effect of viscosity on the hydrodynamic flow~\cite{Dusling:2007gi} builds up throughout the expansion, and one expects the resulting contribution to $w_n^{(\eta,\zeta)}(T)$ to be smooth.
On the other hand, the viscous correction to the momentum distribution at freeze-out~\cite{Teaney:2003kp,Dusling:2009df,Bozek:2009dw} only involves the viscosity at $T_f$. 
Therefore, we decompose $w_n^{(\eta,\zeta)}(T)$ as the sum of a smooth function, which we approximate by a rational function (Pad\'e approximant), and a discrete contribution in the form of a Dirac peak at $T_f$:
\begin{equation}
  \label{deltaplussmooth}
w(T)=w_f\delta(T-T_f)+\frac{a_0+a_1 T+a_2 T^2}{1+b_1 T+b_2T^2+b_3 T^3}
\end{equation}
where we have used the shortcut $w(T)$ for $w_n^{(\eta,\zeta)}(T)$.
The parameters $w_f$, $a_i$ and $b_i$ are fitted to the $\Delta_n$ results using Eq.~(\ref{convolution}). 
The fits are shown as lines in panels (c) and (d) of Fig.~\ref{fig:delta}. 
In order to better constrain the relative magnitudes of the discrete and the smooth contributions, we have carried out a few simulations where $T_0$ is lower than the freeze-out temperature $T_f$ (see Fig.~\ref{fig:delta} (a) and (b)). 
In these simulations, the discrete term dominates the viscous correction. 

The smooth parts of the weighting functions $w_n^{(\eta,\zeta)}(T)$ are displayed in Fig.~\ref{fig:weights}.\footnote{
Note that the variation of $w_n^{(\eta,\zeta)}(T)$ as a function of $T$ follows that of $\Delta_n$ in Fig.~\ref{fig:delta} as a function of $T_0$, except for the values of $T_0$ close to the freeze-out temperature. 
This can be understood easily: 
In the limit where $T_0-T_f\gg \sigma$, inserting Eq.~(\ref{peak}) into Eq.~(\ref{convolution}) and  assuming that  $w_n^{(\eta,\zeta)}(T)$  varies little over a temperature range of order $\sigma$, one can approximate $w_n^{(\eta,\zeta)}(T)\simeq w_n^{(\eta,\zeta)}(T_0)$, and one obtains:
\begin{equation}
\label{inversion}
w^{(\eta)}_n(T_0)\simeq \frac{\Delta_n(T_0)}{(\eta/s)_{\max}\sigma\sqrt{2\pi}},
\end{equation}
and a similar formula for bulk viscosity. 
For temperatures close to the freeze-out temperature $T_f=156.5$~MeV, there are differences between the variations of $\Delta_n$ and $w_n$, which are apparent in particular for shear viscosity, where the leftmost point for $\Delta_{2,3}$ goes up, while the variation of $w_{2,3}^{(\eta)}(T)$ is monotonic down to $T_f$. 
The reason is that for these values of $T_0$, part of the Gaussian profile is cut at $T_f$ (see Fig.~\ref{fig:delta} (a) and (b)), resulting in a smaller value of $\Delta_n$.} 
For shear viscosity, the lowest values of $T$ get the largest weights in absolute value. 
This explains the conclusion from a recent Bayesian study that $\eta/s$ is best constrained in the range $150<T<220$~MeV~\cite{Auvinen:2020mpc}. 
For bulk viscosity, on the other hand, the weight has a peak for intermediate values of the temperature, around 230~MeV for $v_2$, and 190~MeV for $v_3$. 
%%%%%%%%%%%%%%%%%%%%%%%%%%%%%%%%%%%%%%%%%%%%%%%%%%%%%%%%%%%%%%%
\begin{table}
\begin{tabular}{|c|c|r|c|}
\hline
&$n$&$w_f$& $W\equiv\int_{T_f^-} w(T)dT$\cr
\hline
shear&$2$& $-0.07$ & $-1.34$\cr
bulk&$2$& $0.15$ &  $-1.30$\cr
shear&$3$& $-0.49$ & $-2.33$\cr
bulk&$3$& $0.21$ & $-2.61$\cr
\hline
\end{tabular}
\caption{\label{tableweights} 
  Values of $w_f$ (Eq.~(\ref{deltaplussmooth})) and $W$ (Eq.~(\ref{integral})) 
for elliptic ($n=2$) and triangular ($n=3$) flows, and for shear and bulk viscosity, in central Pb+Pb collisions at $\sqrt{s_{\rm NN}}=5.02$~TeV.}
\end{table}
The discrete part $w_f$ of the viscous correction, corresponding to the first term in Eq.~(\ref{deltaplussmooth}), is given in Table~\ref{tableweights}. 
It originates from the viscous correction to the thermal momentum distribution~\cite{Teaney:2003kp}. 
This correction depends on the microscopic dynamics at freeze-out~\cite{Dusling:2009df}, which is not well understood.  
By constrast, the smooth part of Eq.~(\ref{deltaplussmooth}), which is the viscous correction that builds up during the hydrodynamic evolution, solely involves the equations of hydrodynamics and is more robust. 
Looking at the numbers in Table~\ref{tableweights}, one sees that $w_f$ is a small fraction of the integral $W$, which implies that freeze-out only accounts for a small fraction of the viscous suppression: 5\% for $v_2$, 21\% for $v_3$, in the case of a constant $\eta/s$. 
Note also that the bulk viscosity gives a small but {\it positive\/} contribution to $v_n$ at freeze-out.\footnote{
This can be inferred from the observation that $\Delta_n>0$ for $T_0<T_f$ in Fig.~\ref{fig:delta} (d).}
The fact that $w_f$ is  small guarantees that the determination of $\Delta_n$ in viscous hydrodynamics is fairly robust with respect to model uncertainties. 
Note that this is because we have evaluated the $p_t$-integrated $v_n$, which is largely determined by the energy-momentum tensor. 
In a specific $p_t$ range, the sensitivity to  freeze-out dynamics would typically be larger, which is the reason why we do not study $v_n(p_t)$ in this paper. 
As we shall see in Sec.~\ref{s:rhic}, $w_f$ represent a much larger fraction of the viscous correction at lower energies, even for the integrated $v_n$.  

\begin{figure*}[ht]
\begin{center}
\includegraphics[width=\linewidth]{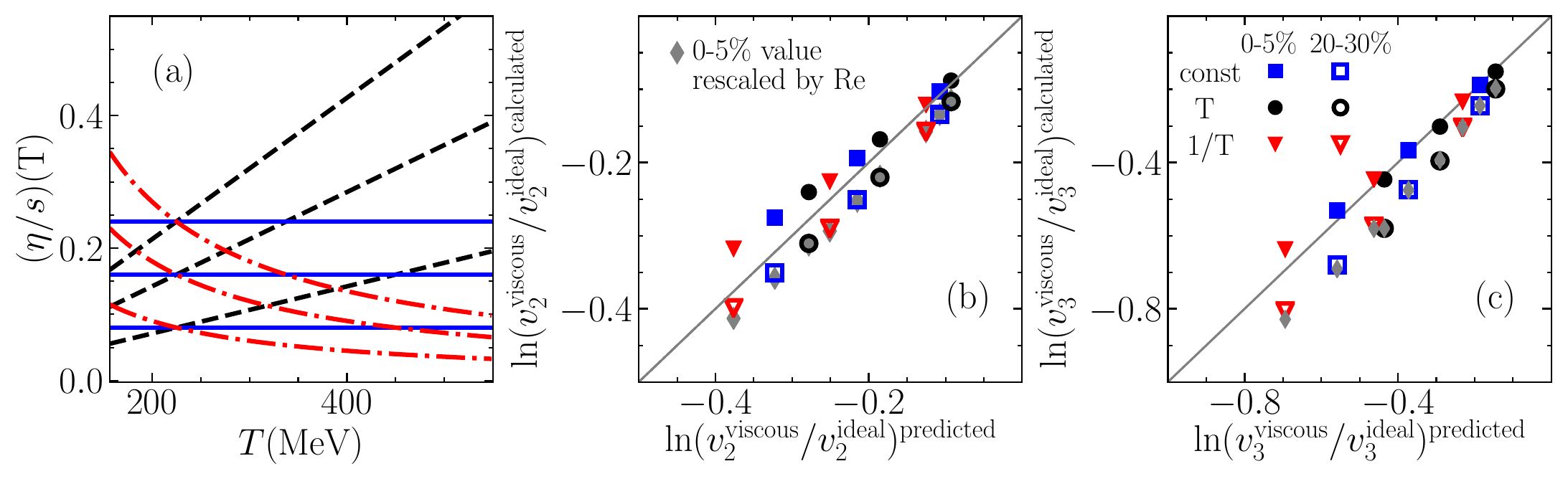} 
\end{center}
\caption{(Color online) 
  (a) Various $(\eta/s)(T)$ profiles which are used to test Eq.~(\ref{convolution}).
They are defined by $(\eta/s)(T)=C$ (full lines), $(\eta/s)(T)=C T/T_0$ (dashed lines) or $(\eta/s)(T)=CT_0/T$, with $T_0=225$~MeV, and $C=0.08$, $0.16$, or $0.24$. 
  Panels (b) and (c) display $\Delta_n$ defined by Eq.~(\ref{defDelta}), as a function of $\Delta_n$ defined by Eq.~(\ref{predicted}).
  Each symbol (squares, circles, triangles) in panels (b) and (c) corresponds to one of the profiles in panel (a). 
  Full symbols correspond to the $0-5\%$ centrality window (Sec.~\ref{s:linearitytests}), and full lines are the diagonals $y=x$. 
  Grey diamonds represent the prediction for $20-30\%$ centrality from Reynolds number scaling, defined by Eq.~(\ref{greysymbols}). 
  Open symbols correspond to the value calculated numerically in the $20-30\%$ centrality window (Sec.~\ref{s:centrality}). 
  \label{fig:others}
}
\end{figure*}    

\begin{figure*}[ht]
\begin{center}
\includegraphics[width=\linewidth]{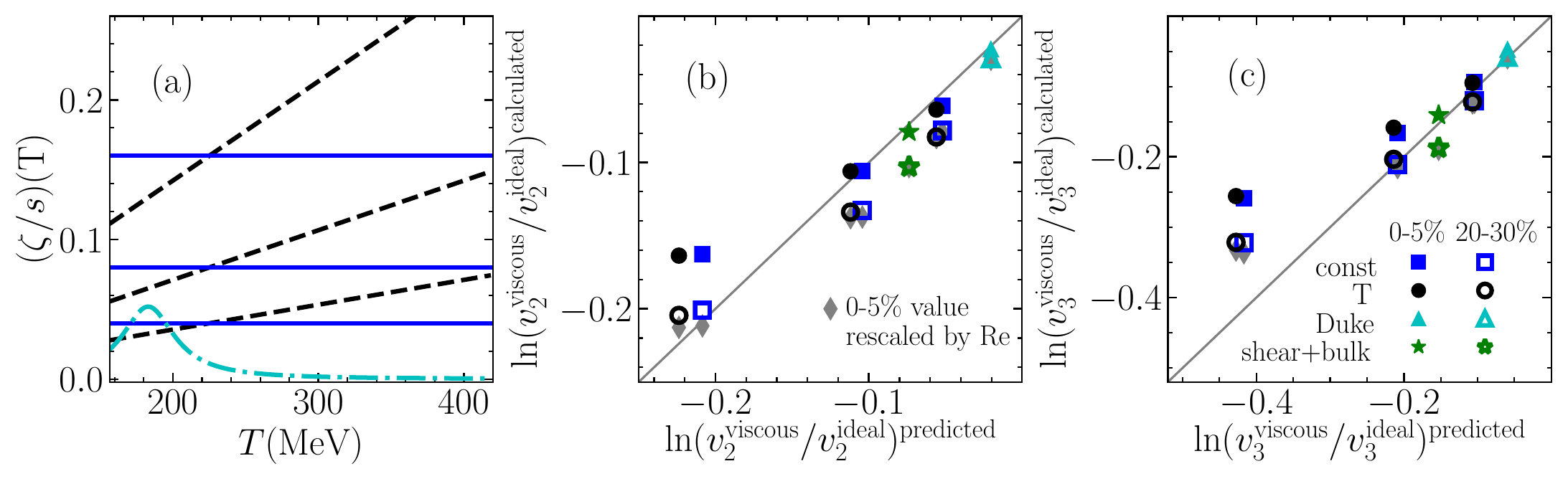} 
\end{center}
\caption{(Color online)
  Same as Fig.~\ref{fig:others} for bulk viscosity alone, and shear+bulk. 
  The $(\zeta/s)(T)$ profiles are represented in panel (a).
  They are defined by $(\zeta/s)(T)=C$ (full lines), $(\zeta/s)(T)=C T/T_0$ (dashed lines), with $T_0=225$~MeV, and $C=0.04$, $0.08$, or $0.16$.
  The dash-dotted line is the Duke parametrization~\cite{Bernhard:2016tnd}. 
  Squares, circles and triangles in panels (b) and (c) correspond to the profiles in panel (a). 
  The stars represent a calculation done with a constant shear viscosity over entropy $\eta/s=0.04$ (we choose a small value so that nonlinear effects are negligible), on top of the Duke parametrization of bulk viscosity.  
  As in  Fig.~\ref{fig:others}, grey diamonds represent the prediction for $20-30\%$ centrality from Reynolds number scaling, defined by Eq.~(\ref{greysymbols}). 
  \label{fig:othersbulk}
}
\end{figure*}    

\section{Orders of magnitude and dimensional analysis}
\label{s:dimensional}

Before we embark on quantitative tests of the ``effective viscosity'' approach, we analyze the order of magnitude of the viscous suppression.
For a constant shear viscosity over entropy ratio $\eta/s=0.08$~\cite{Kovtun:2004de}, Eq.~(\ref{predicted}) together with the numerical values in Table~\ref{tableweights} gives $\Delta_2=-0.11$ and $\Delta_3=-0.19$, corresponding to 10\% and 17\% reductions in $v_2$ and $v_3$, respectively, according to Eq.~(\ref{defDelta}). 

We now check that these numbers are compatible with expectations from dimensional analysis. 
The inverse Reynolds number ${\rm Re}^{-1}$ governs the magnitude of viscous effects. 
It is defined as the ratio of the viscous force, which is $\eta\Delta v$ for shear viscosity, to the inertia, which is $(\epsilon+P)dv/dt$ for a relativistic fluid. 
Assuming that space-time derivatives are of order $1/R$, where $R$ is the rms radius of the initial density profile, and using $\epsilon+P=Ts$, one obtains:
\begin{equation}
\label{reynolds}
      {\rm Re}^{-1}=\frac{(\eta/s)}{T R}.
\end{equation}
In this equation, is it natural to replace $(\eta/s)$ by the effective viscosity $(\eta/s)_{\rm eff}$.
$T$ should be a typical temperature at which viscous effects operate, that is $T\sim 200$~MeV.

In the specific case of anisotropic flow, one can guess the order of magnitude of $\Delta_n$ with the guidance of exact solutions~\cite{Gubser:2010ui}, which give an extra factor of $n^2$~\cite{Teaney:2012ke}.  
The dimensional analysis is the same for bulk viscosity. 
Hence, the back-of-the-envelope estimate of $\Delta_n$ is 
\begin{equation}
  \label{orderofmagnitude}
  \Delta_n
  \sim -n^2 \frac{(\eta/s)_{\rm eff}+(\zeta/s)_{\rm eff}}{T R}.
\end{equation}
Comparing with Eq.~(\ref{predicted}), the expected order of magnitude of the prefactor $W_n^{(\eta,\zeta)}$ is:
\begin{equation}
\label{orderW}
  W_n^{(\eta,\zeta)}\sim -n^2\frac{1}{T R}. 
\end{equation}
With the value $R\simeq 4.2$~fm of our initial condition (Fig.~\ref{fig:density} (a)) and $T\sim 200$~MeV$\simeq 1$~fm$^{-1}$, Eq.~(\ref{orderW}) gives $W_2^{(\eta)}\sim W_2^{(\zeta)}\sim -1.0$ and $W_3^{(\eta)}\sim W_3^{(\zeta)}\sim -2.1$. 
The numerical values in Table~\ref{tableweights} are of the expected order of magnitude. 
In particular, they confirm the expectation that shear and bulk viscosity have similar effects, and that the damping is stronger by a factor $\sim 2$ for $v_3$ than for $v_2$. 

\section{Effective viscosity as a predictor of the damping of $v_n$}
\label{s:linearitytests}

We now test the hypothesis that the effective viscosities (\ref{defeffective}) are good predictors of the viscous suppression $\Delta_n$. 
Using the weights $w_n^{(\eta,\zeta)}(T)$ determined in Sec.~\ref{s:weights}, we can evaluate the effective shear and bulk viscosities for any temperature-dependent viscosity, and then predict the value of the viscous damping $\Delta_n$ using Eq.~(\ref{predicted}).  
In order to check the validity of this prediction, we carry out viscous hydrodynamic simulations with nine different $(\eta/s)(T)$ profiles, which are represented in Fig.~\ref{fig:others} (a),
and seven different $(\zeta/s)(T)$ profiles, which are represented in Fig.~\ref{fig:othersbulk} (a).
These profiles span a wide range of possibilities concerning the variation and magnitude of $\eta/s$ and $\zeta/s$. 

For each of these profiles, panels (b) and (c) of Figs.~\ref{fig:others} and \ref{fig:othersbulk} display the value of $\Delta_2$ and $\Delta_3$ computed numerically in viscous hydrodynamics using Eq.~(\ref{defDelta}), as a function of the value predicted using Eq.~(\ref{predicted}).
When only bulk or shear viscosity is present, the quantity on the $x$ axis is the effective viscosity $(\eta/s)_{n,\rm eff}$ (Fig.~\ref{fig:others}) or  $(\zeta/s)_{n,\rm eff}$ (Fig.~\ref{fig:othersbulk}), multiplied by the corresponding constant $W^{(\eta,\zeta)}_n$.
Note that the effective viscosity is not strictly identical for $n=2$ and $n=3$, because the weights for $n=2$ and $n=3$ in Fig.~\ref{fig:weights} are not exactly proportional to each other.
For a smooth temperature dependence, the effective viscosities associated with $v_2$ and $v_3$ differ little: $(\eta/s)_{2,\rm eff}=1.10 (\eta/s)_{3,\rm eff}$ for $\eta/s\propto T$, $(\eta/s)_{2,\rm eff}=0.94 (\eta/s)_{3,\rm eff}$ for $\eta/s\propto 1/T$ and $(\zeta/s)_{2,\rm eff}=1.05 (\zeta/s)_{3,\rm eff}$ for $\zeta/s\propto T$. 
In the case of the Duke parametrization of the bulk viscosity, which varies quickly precisely in the region where the weights 
$w_n^{(\zeta)}(T)$ vary steeply, the difference is larger:  $(\zeta/s)_{2,\rm eff}=0.69 (\zeta/s)_{3,\rm eff}$. 
 
For small $|\Delta_n|$, the calculated value agrees with the predicted value in all cases:
With only shear viscosity (full symbols in Fig.~\ref{fig:others}), only bulk viscosity (full squares and circles in Fig.~\ref{fig:othersbulk}), or with shear and bulk viscosity simultaneously (stars in Fig.~\ref{fig:othersbulk}). 
This means that Eq.~(\ref{convolution}) holds in the limit of small viscosity, which is precisely the assumption under which it was derived. 
In particular, our calculation shows explicitly that shear viscosity and bulk viscosity give additive contributions to the damping of $v_n$. 

For larger values of $|\Delta_n|$, corresponding to larger values of $\eta/s$, the calculated values (full symbols) start to deviate from the predicted values (full lines). 
They are above, which implies that the dependence of $v_n$ on $(\eta/s)_{n,\rm eff}$ or $(\zeta/s)_{n,\rm eff}$ is slower than exponential.
These nonlinearities are stronger for bulk viscosity than for shear viscosity. 
Despite these deviations, all full symbols almost lie on the same curve. 
This means that the effective viscosity is an excellent predictor of $\Delta_n$, even when viscosity suppresses $v_n$ by a factor 2. 

We now discuss the compatibility of our results with those of Niemi {\it et al.\/}~\cite{Niemi:2015qia}. 
They have carried out extensive simulations with different $\eta/s(T)$ profiles, which have been chosen in such a way that they yields similar $v_2$ and $v_3$. 
We therefore expect that these profiles correspond to similar effective viscosities.
Since, furthermore, one of the profiles is a constant $\eta/s=0.2$, we expect that $(\eta/s)_{n,\rm eff}\sim 0.2$ for all the other profiles, both for $n=2$ and $n=3$.
The authors provide the parameterization for three of these profiles, which are named ``param1'', ``param2'' and ``param4''.
Comparison with our results is not straightforward because they implement partial chemical equilibrium (PCE), and run hydrodynamics down to $T_f=100$~MeV. 
The energy density at $T=100$~MeV with PCE is approximately the same as at $T=140$~MeV without PCE.
We try to take this difference into account, at least approximately, by evaluating the effective viscosity (\ref{defeffective}) with a lower value $T_f=140$~MeV. 
For this purpose, we extrapolate the weights $w_n^{(\eta)}$ down to 140~MeV using Eq.~(\ref{deltaplussmooth}), and we neglect the discrete contribution, which is small but depends on $T_f$, since we have not evaluated $w_f$ for $T_f=140$~MeV.
We obtain $(\eta/s)_{2,{\rm eff}}=0.175$ and $(\eta/s)_{3,{\rm eff}}=0.163$ for ``param1'', $0.184$ and $0.185$ for ``param2'', and  $0.206$ and $0.207$ for ``param4''. 
As expected, all effective viscosities are close to $0.2$.
The ordering explains the fine splitting observed in Fig.~14 (a) of Ref.~\cite{Niemi:2015qia}, which shows that the damping is weakest for the ``param1'' parametrization, and strongest for the ``param4'' parametrization. 
This shows that effective viscosities can used to efficiently classify parametrizations of temperature-dependent viscosities.

\section{Centrality and system-size dependence}
\label{s:centrality}

We show that at a given collision energy, the dependence of $\Delta_n$ on nuclear size and collision centrality is determined by the $1/R$ dependence expected from Reynolds number scaling, Eq.~(\ref{orderofmagnitude}), and that the effective viscosity is unchanged.
We first present the general argument, then the numerical results that support it.

The key observation is that the mean transverse momentum of outgoing hadrons, $\langle p_t\rangle$, is almost independent of centrality and system size:
Specifically, $\langle p_t\rangle$ varies by less than 1\% between 0 and 30\% centrality in Pb+Pb collisions at 
$5.02$~TeV~\cite{Acharya:2018eaq}, while the multiplicity decreases by a factor $\sim 3$~\cite{Adam:2015ptt}.
$\langle p_t\rangle$ also differs by less than 2\% in Pb+Pb and Xe+Xe collisions~\cite{Acharya:2018eaq}, while the multiplicity changes by a factor $\sim 1.6$. 

In ideal hydrodynamics, the mean transverse momentum is unchanged under a uniform scaling of space-time coordinates, where the entropy density $s$ and the fluid velocity $u^\nu$ are unchanged:
\begin{eqnarray}
  \label{scale}
  x^\mu &\rightarrow& \lambda x^\mu\cr
  s(x^\mu)&\rightarrow& s(\lambda x^\mu)\cr
  u^\nu(x^\mu)&\rightarrow& u^\nu(\lambda x^\mu)
\end{eqnarray}
The volume and the final multiplicity, which are extensive quantities, are multiplied by $\lambda^3$, but 
$\langle p_t\rangle$, which is an intensive quantity, remains the same.
Reversing the argument, the observation that the mean transverse momentum remain constant as one varies centrality or system size implies that these variations amount, to a good approximation, to a uniform scaling~\cite{Gardim:2019xjs}.
A less central collision, or a smaller nucleus, goes along with a faster expansion, but with the same density and temperature. 
This statement may seem counter intuitive, as one would think that more central collisions or larger nuclei imply a higher density.
However, one typically has in mind a comparison at the same time, while the time coordinate should also be rescaled in Eq.~(\ref{scale}): 
One should evaluate the density at an earlier time for the smaller system. 
A uniform scaling does not change the fraction of the space-time history that the system spends at a given temperature.
Therefore, the effective viscosity, which represents the relative weights of the different temperatures, is unchanged. 

The observation that the density is essentially constant as a function of centrality or system size is supported by the following theoretical argument:  
The multiplicity is approximately proportional to the number of constituent quarks~\cite{Eremin:2003qn}, which is also proportional to the volume since the nuclear density is approximately constant. 
Hence the ratio multiplicity/volume does not vary significantly. 
Note that the shape changes as a function of the collision centrality (see Fig.~\ref{fig:density}), so that the scaling is not strictly isotropic.
The anisotropy is responsible for anisotropic flow, but has a small effect on the mean transverse momentum. 

We simulate mid-central collisions by adjusting the parameters of the initial density profile (\ref{initial}) in our hydrodynamic calculation (Fig.~\ref{fig:density} (b)). 
Note that we keep the same value of the initial time ($\tau_0=0.6$~fm/c) for both centralities, while it should also be rescaled for the transformation (\ref{scale}) to be exact. 
However, this breaking of scale invariance occurs long before $v_n$ develops, and we will see that it has no effect on the final results.  
In order to ensure that $\langle p_t\rangle$ is the same for both centralities, we require that, at a time proportional to the the rms radius $R$, the entropy density is the same.
We therefore choose $s_0$ in such a way that $s_0\tau_0/R$ is unchanged.
$s_0$ and $R_0$ are finally fixed by requesting that the final multiplicity matches the experimental value, which yields the profile in Fig.~\ref{fig:density} (b).\footnote{Due to the anisotropies, the rms radius $R$ does not coincide with the parameter $R_0$ in Eq.~(\ref{initial}). For the profile in Fig.~\ref{fig:density} (b), for instance, $R_0=2.97$~fm, while $R=3.19$~fm.} 

\begin{figure*}[ht]
\begin{center}
\includegraphics[width=.8\linewidth]{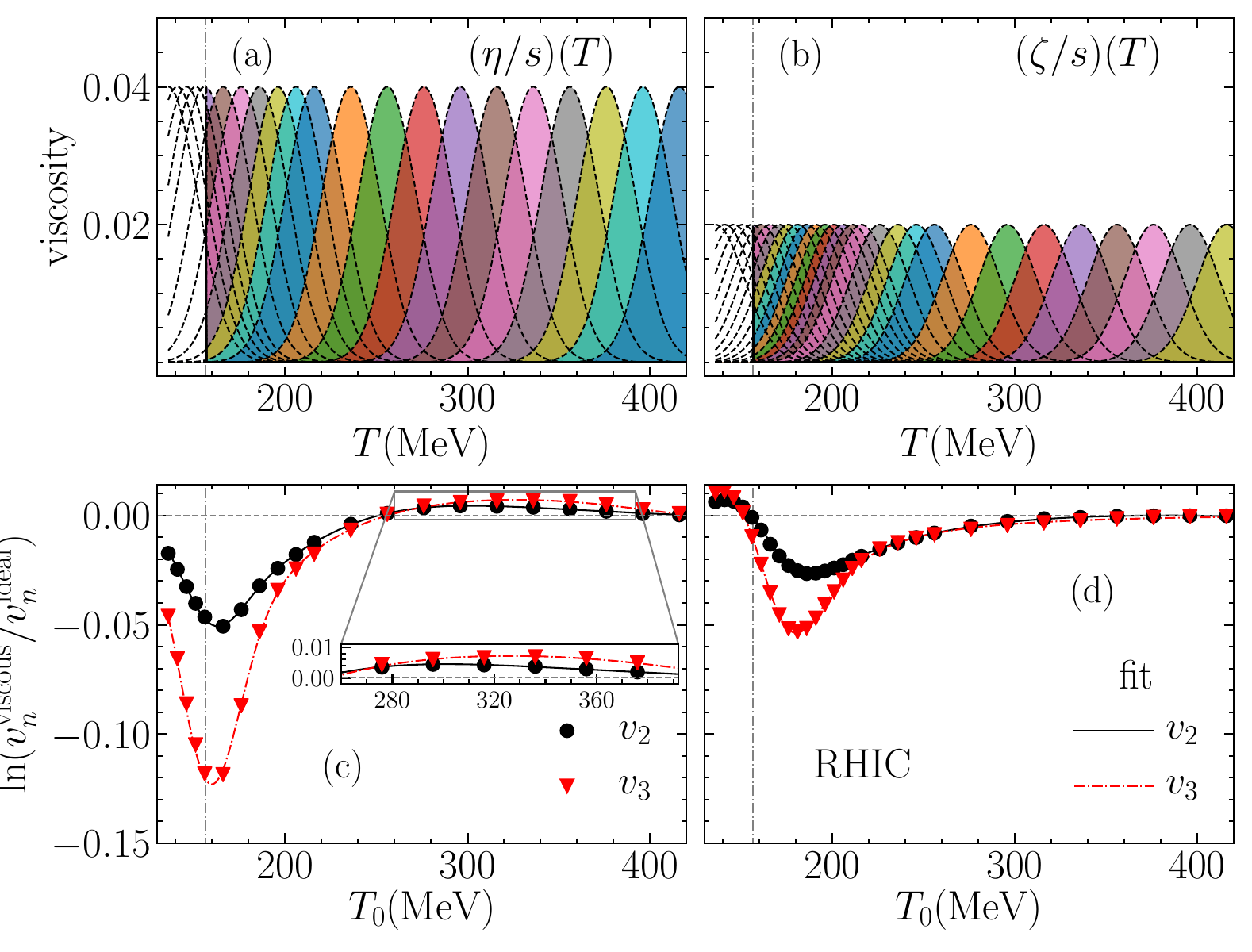} 
\end{center}
\caption{(Color online)
  Same as Fig.~\ref{fig:delta}, for the collision energy $\sqrt{s_{\rm NN}}=200$~GeV.
  }
\label{fig:deltarhic}
\end{figure*}

\begin{figure}[ht]
\begin{center}
\includegraphics[width=\linewidth]{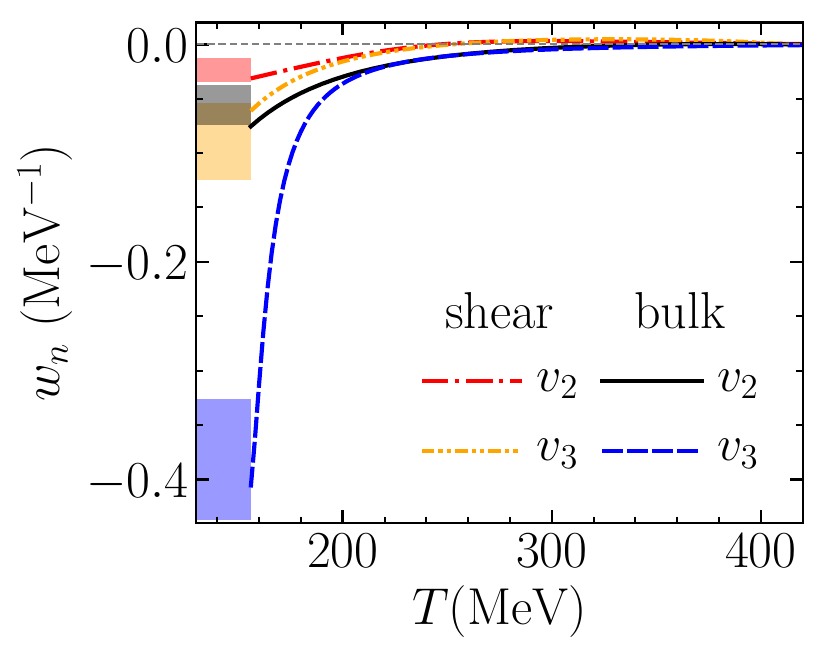} 
\end{center}
\caption{(Color online)
Same as Fig.~\ref{fig:weights} for $\sqrt{s_{\rm NN}}=200$~GeV. 
}
\label{fig:weightsrhic}
\end{figure}    

We then evaluate the viscous suppression of $v_n$ in the 20-30\% centrality range for the same temperature-dependent viscosities as in Sec.~\ref{s:linearitytests}. 
Results are displayed in Figs.~\ref{fig:others} and \ref{fig:othersbulk} as open symbols.
We plot the value of $\Delta_n$ calculated numerically with (\ref{defDelta}), as a function of the value calculated for central collisions. 
Therefore, closed and open symbols, corresponding to central and mid-central collisions, are vertically aligned.
The open symbols are below the closed symbols, which means that the viscous suppression is larger for mid-central than for central collisions. 
The grey diamonds represent the prediction from dimensional analysis, that $\Delta_n$ is proportional to $1/R$ (Eq.~(\ref{orderofmagnitude})):
\begin{eqnarray}
\label{greysymbols}
\Delta_n[20-30\%]&=&\frac{R[0-5\%]}{R[20-30\%]}\Delta_n[0-5\%]\cr&=&1.32\, \Delta_n[0-5\%].
\end{eqnarray}
Grey diamonds lie almost on top of open symbols, which confirms that the centrality dependence of the viscous suppression is determined by Reynolds number scaling. 
The explicit calculation above is only done for one collision system and one centrality range, but conclusions are general.
As long as $\langle p_t\rangle$ does not vary, the effective viscosity should remain the same, and the centrality dependence of the Reynolds number is solely determined by the transverse size.  
The decrease of $\langle p_t\rangle$ becomes significant for peripheral collisions, but this is also the place where the hydrodynamic description is less reliable. 

\section{Effective viscosities at RHIC}
\label{s:rhic}

While the effective viscosity is roughly independent of system size and centrality, it depends on the collision energy.
The lower the collision energy, the lower the temperature of the quark-gluon matter formed in the collision~\cite{Gardim:2019xjs}, and one expects this change to reflect on the weights $w_n^{(\eta,\zeta)}(T)$ entering the effective viscosities (\ref{defeffective}).
In order to illustrate this dependence on collision energy, we carry out simulations at the top RHIC energy $\sqrt{s_{\rm NN}}=200$~GeV.
We single out the dependence on collision energy by changing only the normalization constant $s_0$ in the initial density profile (\ref{initial}), and keeping all other parameters ($R_0$, $\varepsilon_2$, $\varepsilon_3$) constant. 
For ideal hydrodynamics, we choose $s_0=173$~fm$^{-3}$, which ensures that the charged multiplicity per nucleon matches the value measured in central Au+Au collisions~\cite{Back:2002uc}.

%%%%%%%%%%%%%%%%%%%%%%%%%%%%%%%%%%%%%%%%%%%%%%%%%%%%%%%%%%%%%%%
\begin{table}
\begin{tabular}{|c|c|r|c|}
\hline
&$n$&$w_f$& $W\equiv\int_{T_f^-} w(T)dT$\cr
\hline
shear&$2$& $-0.67$ & $-1.56$\cr
bulk&$2$& $1.12$ &  $-2.33$\cr
shear&$3$& $-2.12$ & $-3.40$\cr
bulk&$3$& $3.34$ & $-3.81$\cr
\hline
\end{tabular}
\caption{\label{tableweightsrhic} 
Same as Table~\ref{tableweights} for $\sqrt{s_{\rm NN}}=200$~GeV.}
\end{table}

The calculation is done exactly as in Sec.~\ref{s:weights}. 
Fig.~\ref{fig:deltarhic} is the equivalent of Fig.~\ref{fig:delta}, but at the lower energy, where the temperature is $\sim 25\%$ smaller. 
Results for $\Delta_n$ are comparable, except for the overall temperature scale, and the overall magnitude. 
In particular, the small increase of $v_n$ due to shear viscosity, which was observed for $T_0>330$~MeV in Fig.~\ref{fig:delta} (c), is now observed for $T_0>250$~MeV (see the zoom in Fig.~\ref{fig:deltarhic} (c)). 

The $\Delta_n$ results are then fitted using Eqs.~(\ref{convolution}) and (\ref{deltaplussmooth}). 
The integral of the weights $W_n^{(\eta,\zeta)}$, which determine the damping for a given effective viscosity according to 
Eq.~(\ref{predicted}), are larger at RHIC (Table~\ref{tableweightsrhic}) than at the LHC (Table~\ref{tableweights}). 
This implies that for a given effective viscosity, the damping is somewhat stronger at RHIC than at LHC.
This increase is a natural consequence of the lower temperature, as shown by the estimate (\ref{orderW}) from dimensional analysis. 
One would expect an increase by a factor $\sim 1.3$, but the increase is significantly larger, by a factor $\sim 1.8$, for the effect on bulk viscosity on $v_2$. 
This is probably a consequence of the non-conformal equation of state. 
RHIC probes the equation of state just above the transition to the quark-gluon plasma, and going from RHIC to LHC does not boil down to  rescaling the temperature. 
The most spectacular difference between RHIC and LHC results is that the discrete contribution $w_f$, corresponding to the viscous correction at freeze out, is now a sizable fraction of the total $W$ (Table~\ref{tableweightsrhic}). 
Since, as pointed out in Sec.~\ref{s:weights}, there is a theoretical uncertainty on $w_f$, it implies that the determination of the effective viscosity is less robust at RHIC than at the LHC.

The smooth part of the weights, displayed in Fig.~\ref{fig:weightsrhic}, differs significantly from the result at the higher energy (Fig.~\ref{fig:weights}), in particular for bulk viscosity.
This is somewhat surprising as the results in Fig.~\ref{fig:deltarhic}, from which they are obtained, are similar to the results in Fig.~\ref{fig:delta}, from which Fig.~\ref{fig:weights} is obtained.
The difference is likely due to the separation between the discrete and the smooth contribution.
In particular, the large negative value of $w_3^{(\zeta)}$ just above the freeze-out temperature partially compensates the effect of the large positive contribution {\it at\/} freeze out. 

\section{Conclusions}
Within the hydrodynamic description of heavy-ion collisions, we have evaluated the dependence of elliptic and triangular flows on shear and bulk viscosities, for an arbitrary temperature dependence of these transport coefficients. 
We have assumed that $v_2$ and $v_3$ are determined by linear response to the initial anisotropies $\varepsilon_2$ and $\varepsilon_3$, and studied the dependence of the response on viscosity, thereby generalizing the study of Teaney and Yan~\cite{Teaney:2012ke}, which was done for a constant $\eta/s$. 
We have shown that the damping is the sum of contributions from shear and bulk viscosity.
Each of these contributions is determined by effective shear and bulk viscosities, which are weighted averages of the temperature-dependent viscosities.

The effective viscosities consist of a discrete part, proportional to the viscosity at freeze out, and a continuous part, which is a weighted integral of the viscosity over temperatures above the freeze-out temperature.
The discrete part originates from the off-equilibrium correction to the momentum distribution of outgoing particles, while the continuous part is due to the hydrodynamic expansion itself. 
For the integrated $v_n$, the discrete part is a small contribution at LHC energies.
This guarantees that the determination of $v_n$ in viscous hydrodynamics is robust with respect to uncertainties on the theoretical description of the hadronic phase. 
At RHIC energies, on the other hand, the discrete and the continuous contributions to the effective viscosities are of the same order of magnitude, which entails a large theoretical uncertainty. 

The weights defining the effective viscosities are displayed in Figs.~\ref{fig:weights} and~\ref{fig:weightsrhic}. 
Shear viscosity matters in the range $T<280$~MeV at the LHC, $T<210$~MeV at RHIC.
For bulk viscosity, the weights decrease less quickly, so that higher temperatures, corresponding to earlier stages of the expansion, are comparatively more important.

We have shown that the effective viscosity is independent of centrality and system size at a given collision energy, to the same extent as the mean transverse momentum $\langle p_t\rangle$. 
The dependence of the damping on centrality and system size follows the $1/R$ dependence expected on the basis of Reynolds number scaling, where $R$ is the transverse radius. 
Furthermore, effective viscosities are very similar for $v_2$ and $v_3$, which implies that a combined analysis of all existing $v_2$ and $v_3$ data at a given energy can at best constrain two numbers: the effective shear and bulk viscosities at this energy.
This in turn implies that the temperature dependence of transport coefficients cannot be extracted from LHC data alone, and claims from early Bayesian analyses~\cite{Bernhard:2019bmu} must be revisited carefully~\cite{Paquet:2020rxl}.\footnote{Note that the extraction of the effective viscosity from data relies crucially also involves the detailed modeling of initial conditions~\cite{Luzum:2008cw,Retinskaya:2013gca}, which is beyond the scope of the present work.} 
Global analyses should be more efficient if they make use of the observation that data at a given energy only give access to the effective viscosity at that energy.
Detailed information about the temperature dependence of transport coefficients can only be obtained by a simultaneous fit to RHIC and LHC data, as recognized by the recent analysis of the JETSCAPE collaboration~\cite{Everett:2020yty}. 
If, for instance, the shear viscosity over entropy ratio was large only above $200$~MeV, damping of anisotropic flow would be larger at the LHC than at RHIC.

\section*{Acknowledgments}
FGG was supported by CNPq (Conselho Nacional de Desenvolvimento Cientifico) grant 312932/2018-9, by  INCT-FNA grant 464898/2014-5 and FAPESP grant 2018/24720-6.
J.-Y.O. was supported by USP-COFECUB (grant Uc Ph 160-16, 2015/13).
We thank Giuliano Giacalone for discussions.

\end{document}